\documentclass[a4paper]{jpconf}
\usepackage{graphicx}
\usepackage{textcomp}
\begin{document}
\title{Jet-hadron correlations in STAR}

\author{Alice Ohlson for the STAR Collaboration}

\address{Physics Department, Yale University, 272 Whitney Ave., New Haven, 06520-8120 CT, USA.}

\ead{alice.ohlson@yale.edu}

\begin{abstract}
In recent years, the study of dihadron correlations has been one of the primary methods used to investigate the propagation and modification of hard-scattered partons through the QGP.  Due to recent advances in jet-finding algorithms, it is now possible to use reconstructed jets in these correlation studies, extending the kinematic reach compared to dihadron analyses.  The results of the jet-hadron correlation analysis indicate a broadening and softening of jets that interact with the medium.  Jet-hadron correlations can also be used to assess the systematics of other jet-like correlation analyses, such as 2+1 correlations.  It is shown that the jets selected in 2+1 correlations are relatively unmodified.  Future work will include an analysis of jet-hadron correlations with respect to the event plane to measure the pathlength dependence of parton energy loss.  The first steps in this analysis indicate that complications arise when calculating the event plane in the presence of a jet as well as in calculating jet $v_2$.  The data analyzed were collected by the STAR detector in $\sqrt{s_{NN}}$ = 200 GeV Au--Au collisions at the Relativistic Heavy Ion Collider (RHIC).
\end{abstract}

\section{Introduction}
Jets are used to study the energy loss of partons which traverse the hot and dense medium produced in ultrarelativistic heavy-ion collisions.  Models of partonic energy loss fall into two general categories: radiative/collisional energy loss models\cite{ref:radcoll} (where radiative energy loss is expected to dominate for light partons) and ``black-and-white'' models (for example, the core-corona model\cite{ref:corecorona}).  In the former, partons lose energy and are scattered as they traverse the medium.  In angular correlation studies this would manifest itself as a difference between the widths and yields of the jet peaks in Au--Au compared to p--p.  In the latter, jets either exit the medium entirely unmodified or are completely thermalized and indistinguishable from the bulk.  While the overall rate of jets would be suppressed in inclusive studies, in angular correlation studies (which are normalized per trigger jet) there would be no observed difference in jet shapes between Au--Au and p--p.  These two classes of models can be distinguished by analyzing jet shapes and yields with jet-hadron angular correlations\cite{ref:HP10}.  

\section{Data sets and Analysis}
In jet-hadron correlations the azimuthal distribution of associated particles (all charged hadrons in the event) is studied with respect to the axis of a reconstructed jet.  Utilizing a reconstructed jet as a trigger increases the range of accessible parton kinematics compared to dihadron correlations\cite{ref:dihadron}.  

The data analyzed here were collected by the STAR detector at the Relativistic Heavy Ion Collider (RHIC) in Run 6 p--p and Run 7 Au--Au collisions at $\sqrt{s_{NN}}$ = 200 GeV.  The main detectors used in this analysis are the Time Projection Chamber (TPC) and the Barrel Electromagnetic Calorimeter (BEMC).  Both detectors have full $2\pi$ azimuthal coverage and span a pseudorapidity range of $|\eta| < 1$.  Charged tracks are reconstructed in the TPC and the BEMC measures the neutral energy component.  Corrections for double-counting of electrons and charged hadronic energy deposition in the BEMC are applied. 

%\subsection{Trigger Jet Reconstruction}
Jets are reconstructed using the anti-$k_{T}$ algorithm from the FastJet package\cite{ref:fj}.  Only tracks and neutral towers with $p_{T}$\textgreater 2 GeV/$c$ are used in the jet reconstruction in order to minimize the effect of background fluctuations.  Furthermore, the reconstructed jets used in this analysis must include a BEMC tower that fired the high tower (HT) trigger.  The online HT trigger requires that more than 5.4 GeV of transverse energy is deposited in one BEMC tower ($\Delta\varphi \times \Delta\eta = 0.05 \times 0.05$).  It is expected that the $p_T$ cut and HT trigger requirements bias the trigger selection towards harder fragmenting jets which interact less with the medium, potentially due to surface production\cite{ref:trenk}.  

\begin{figure}
\begin{centering}
\includegraphics[width=170mm, trim = 25mm 9mm 2cm 7mm, clip=true]{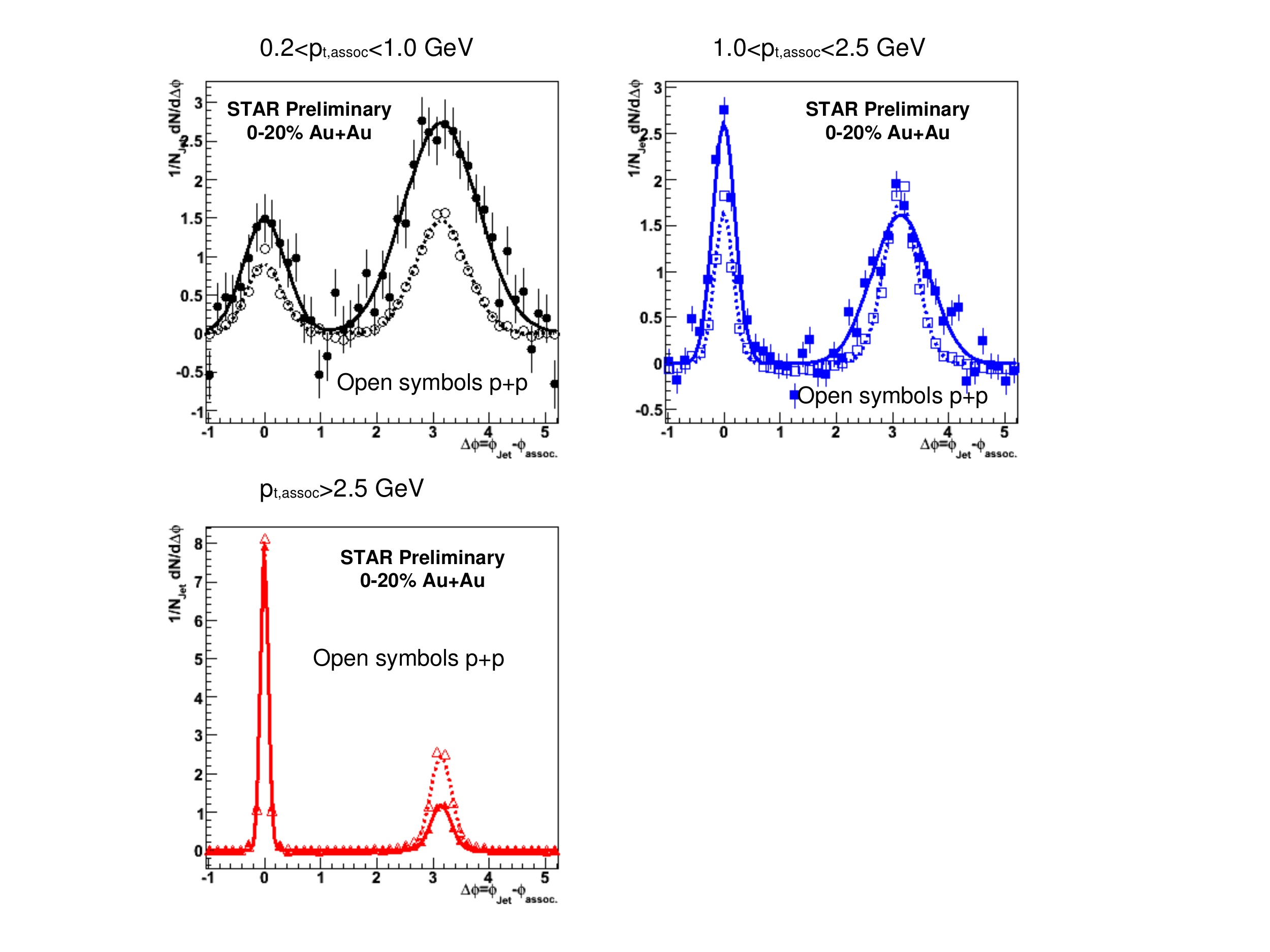}%\hspace{1pc}%
\caption{\label{fig:raw}Jet-hadron correlations with a flat background (no $v_2$ modulation) subtracted for jets with reconstructed $p_T^{jet}$ \textgreater 20 GeV/$c$.  Three $p_T^{assoc}$ bins are shown: top left 0.2 \textless $p_T^{assoc}$ \textless 1.0 GeV/$c$, top right 1.0 \textless $p_T^{assoc}$ \textless 2.5 GeV/$c$, bottom left $p_T^{assoc}$ \textgreater 2.5 GeV/$c$.  Closed symbols are Au--Au (0-20\% centrality) and open symbols are p--p.  }
\end{centering}
\end{figure}

%\subsection{Background Subtraction}
The raw jet-hadron correlations are $\Delta\varphi$ distributions, where $\Delta\varphi = \varphi_{jet} - \varphi_{assoc}$.  In Au--Au the raw correlations include a large contribution from the heavy-ion background.  It is necessary to subtract this background in order to analyze the jet signal (the nearside and awayside jet peaks at $\Delta\varphi = 0$ and $\Delta\varphi = \pi$, respectively).  There are two complications associated with this background subtraction: First, using the ZYAM (``Zero Yield at Minimum'') technique is unfavorable because it overestimates the background level in the presence of broad jet peaks.  Second, the $v_2$ of the trigger jets is \textit{a priori} unknown.  In this analysis background levels are estimated by fitting.  The signal is assumed to have the functional form in Equation \ref{eq:bkgfit}.  
\begin{equation}\label{eq:bkgfit}
\mbox{Nearside Gaussian} + \mbox{Awayside Gaussian} + B\left(1+2v_2^{assoc}v_2^{jet}\cos(2\Delta\varphi)\right)
\end{equation}
In Equation \ref{eq:bkgfit}, $v_2^{assoc}$ is given by the standard STAR parameterization ($^1/_2(v_2\{2\}+v_2\{4\})$ as a function of $p_T$) and $v_2^{jet}$ is estimated to be $v_2\{2\}$ at $p_T = 6\mbox{ GeV}/c$.  Because $v_2^{jet}$ is unknown, the lower and upper bounds on the uncertainty of $v_2^{assoc}v_2^{jet}$ are conservatively estimated to be 0 and 150\% of $v_2\{2\}(6\mbox{ GeV}/c) \cdot v_2\{2\}(p_T^{assoc})$.  The correlations with a flat background subtracted are shown in Figure \ref{fig:raw} for three $p_T^{assoc}$ bins.  The awayside peak appears enhanced in Au--Au compared to p--p in the lowest $p_T^{assoc}$ bin and suppressed in the highest $p_T^{assoc}$ bin.  In the intermediate $p_T^{assoc}$ bin the Au--Au awayside peak appears broader than in p--p.  

\section{Jet-Hadron Correlations - The Nearside}
The highly biased trigger jet sample allows easier comparison between nearside jets in Au--Au and p--p.  On average, background particles with $p_T > 2\mbox{ GeV}/c$ create an excess of $p_T \sim 1\mbox{ GeV}/c$ within the jet cone.  The energy of the p--p reference jets is consequently shifted down by 1 GeV/$c$ with respect to the Au--Au jet energies to more accurately compare similar partonic energies.  When comparing the Gaussian widths of the nearside jet peaks in p--p and Au--Au, the Au--Au peaks appear broader (even after the -1 GeV/$c$ shift).  Additionally, there seems to be slight enhancement in the ratio of the nearside associated yields, $I_{AA}$ (Equation \ref{eq:Iaa}), below $p_T^{assoc} = 2\mbox{ GeV}/c$.  
\begin{equation}\label{eq:Iaa}
I_{AA}(p_{T}^{assoc}) = \frac{Y_{AA}(p_{T}^{assoc})}{Y_{pp}(p_{T}^{assoc})}
\end{equation}
There are two possibilities for this enhancement: (1) The enhancement in the low-$p_T^{assoc}$ yield could be due to the ridge, $v_3$, or other bulk effects that have not been accounted for (i.e. the enhancement is related to the bulk), or (2) it could be due to energy loss on the nearside (i.e. the enhancement is related to the jet).  Under the assumption that the latter possibility is the cause for the enhancement, then the energy loss can be estimated by summing up the excess $p_T$ below $p_T^{assoc} = 2\mbox{ GeV}/c$, yielding a charged energy difference $\Delta E \sim $2 GeV/$c$.  This can be accounted for by adjusting the p--p reference up by $1.5\cdot\Delta E$ (where the factor of 1.5 is due to the neutral energy fraction).  

The possible modification of the nearside jet peak can also be quantified by the difference in energy between Au--Au and p--p as a function of $p_T^{assoc}$, $D_{AA}$ (Equation \ref{eq:Daa}), and its integral, $\Delta B$ (Equation \ref{eq:deltaB}).  
\begin{equation}\label{eq:Daa}
D_{AA}(p_{T}^{assoc}) = Y_{AA}(p_{T}^{assoc}) \cdot p_{T,AA}^{assoc} - Y_{pp}(p_{T}^{assoc}) \cdot p_{T,pp}^{assoc}
\end{equation}
\begin{equation}\label{eq:deltaB}
\Delta B = \int d p_{T}^{assoc} D_{AA}(p_{T}^{assoc})
\end{equation}
When $\Delta B \sim 0$ the suppression at high-$p_T$ is fully balanced by enhancement at low-$p_T$.  On the nearside it is observed that $\Delta B \sim 1-2\mbox{ GeV}/c$ before the p--p energy shift, indicating that the energy is in large part balanced.  After the p--p reference is shifted by $1.5\cdot \Delta E$, the nearside $\Delta B$ is even smaller.  This indicates the possibility that there is energy loss on the nearside, even in high tower trigger jets with stringent jet-finding criteria, although this certainly needs to be studied more thoroughly.  In the following analysis of the awayside jets, Au--Au will be compared to p--p with both energy shifts applied (+3 GeV/$c$ and -1 GeV/$c$).  The uncertainty on the trigger jet energy scale is estimated to be $\pm2$ GeV/$c$.  

\section{Jet-Hadron Correlations - The Awayside}
\begin{figure}
\includegraphics[width=100mm]{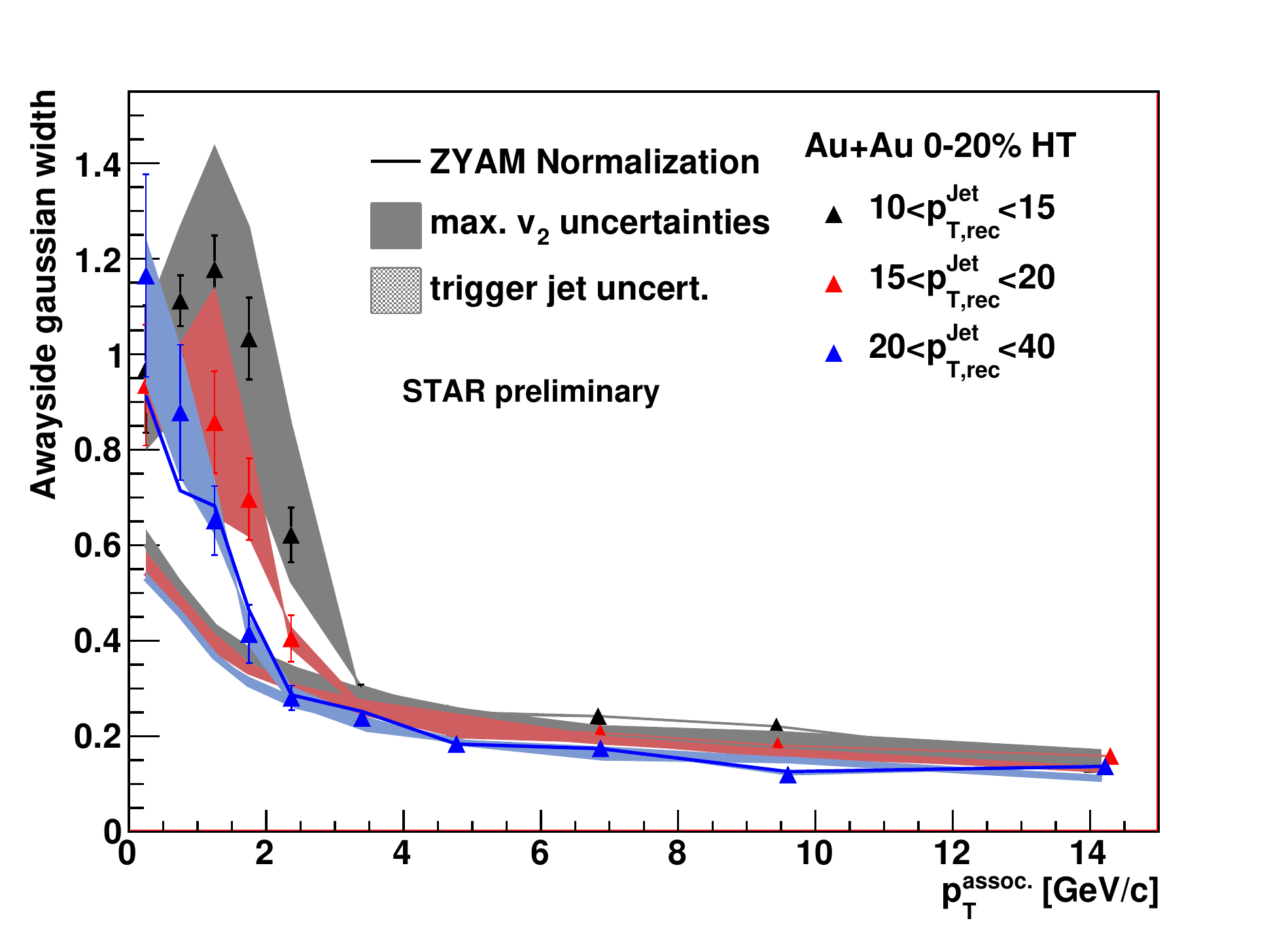}%\hspace{1pc}%
\begin{minipage}[b]{14pc}\caption{\label{fig:asw}The Gaussian widths of the awayside jet peaks in Au--Au (upper shaded bands with symbols) are compared to p--p (lower shaded bands without symbols) for three reconstructed jet energy ranges: 10 \textless $p_T^{jet}$ \textless 15 GeV/$c$ in black, 15 \textless $p_T^{jet}$ \textless 20 GeV/$c$ in red, 20 \textless $p_T^{jet}$ \textless 40 GeV/$c$ in blue.  The awayside jet peaks in Au--Au are significantly broadened compared to p--p for jet energies between 10 and 40 GeV/$c$.}
\end{minipage}
\end{figure}

\begin{figure}
\includegraphics[width=100mm]{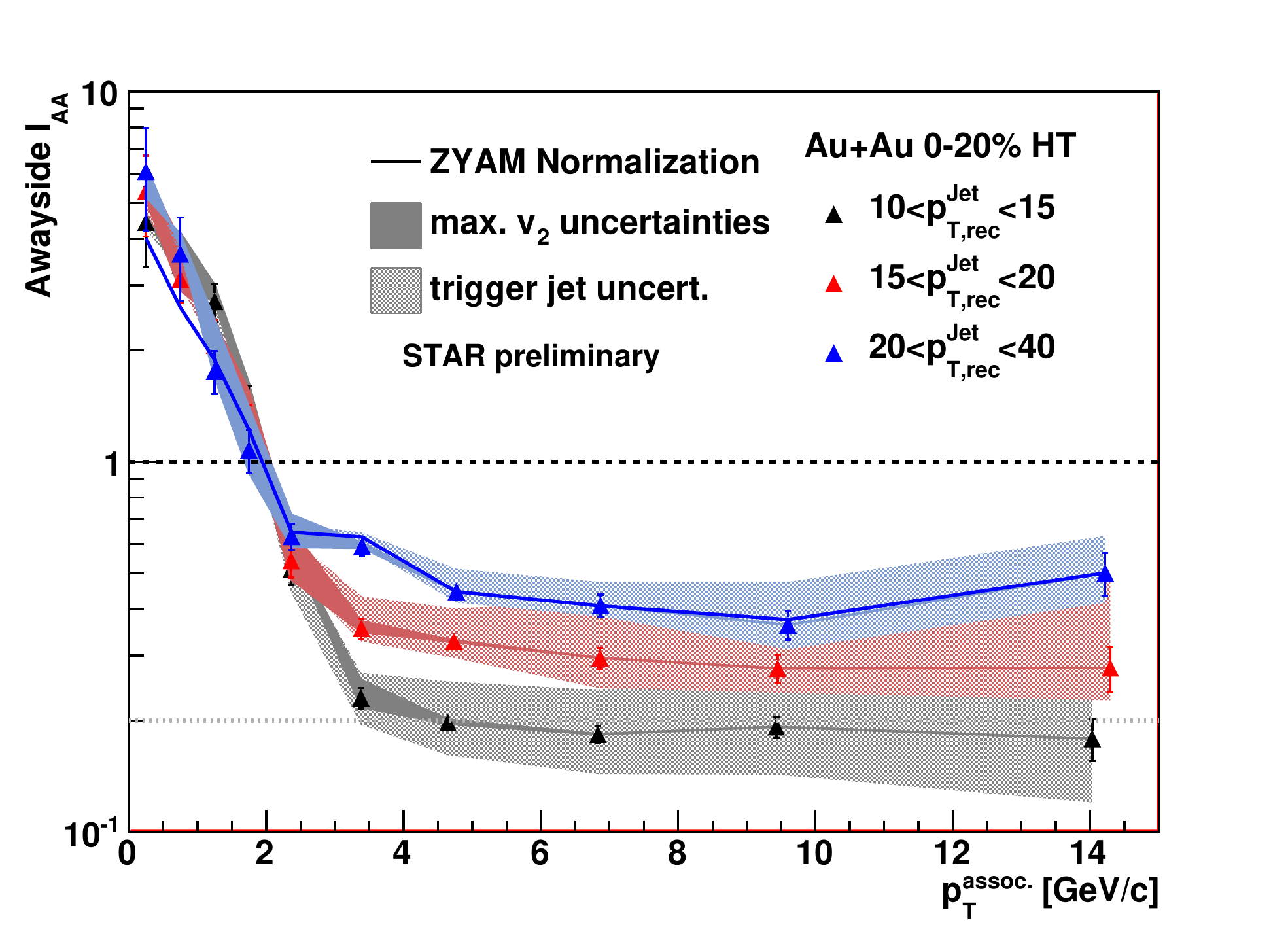}%\hspace{1pc}%
\begin{minipage}[b]{14pc}\caption{\label{fig:asiaa}The awayside $I_{AA}$ (defined in Equation \ref{eq:Iaa}) is shown for three reconstructed jet energy ranges: 10 \textless $p_T^{jet}$ \textless 15 GeV/$c$ in black, 15 \textless $p_T^{jet}$ \textless 20 GeV/$c$ in red, 20 \textless $p_T^{jet}$ \textless 40 GeV/$c$ in blue.  $I_{AA}(p_T^{assoc}) = 1$ would indicate no difference in the awayside jet peak charged hadron yield between Au--Au and p--p.  The awayside $I_{AA}$ indicates significant softening of the awayside jet for jet energies between 10 and 40 GeV/$c$.}
\end{minipage}
\end{figure}

\begin{figure}
\includegraphics[width=100mm]{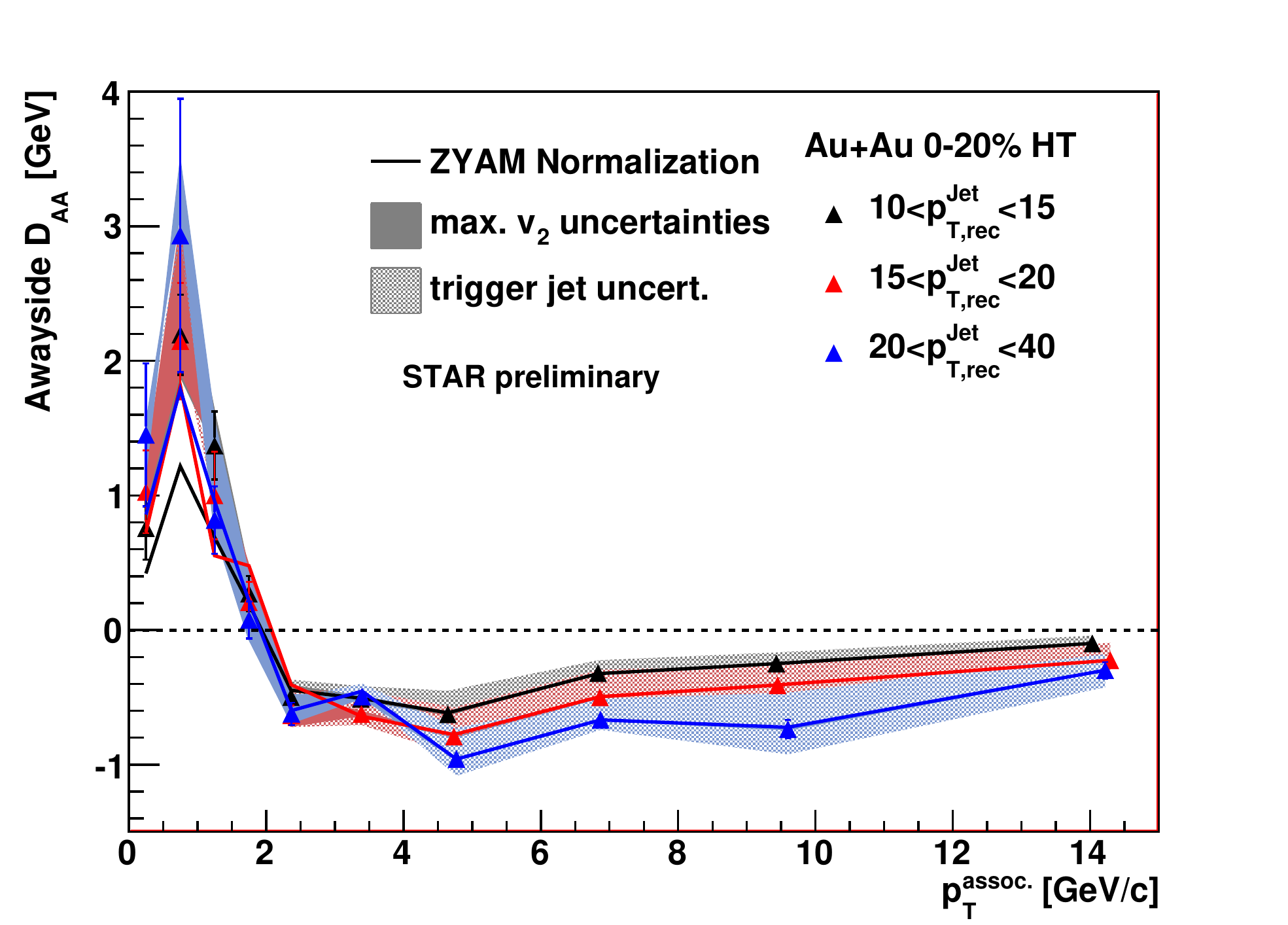}%\hspace{1pc}%
\begin{minipage}[b]{14pc}\caption{\label{fig:asdaa}The awayside $D_{AA}$ (defined in Equation \ref{eq:Daa}) is shown for three reconstructed jet energy ranges: 10 \textless $p_T^{jet}$ \textless 15 GeV/$c$ in black, 15 \textless $p_T^{jet}$ \textless 20 GeV/$c$ in red, 20 \textless $p_T^{jet}$ \textless 40 GeV/$c$ in blue.  $D_{AA}(p_T^{assoc}) = 0$ would indicate no difference in the energy of the awayside jet fragments between Au--Au and p--p.  The awayside $D_{AA}$ illustrates how high-$p_T^{assoc}$ suppression is compensated for by low-$p_T^{assoc}$ enhancement.}
\end{minipage}
\end{figure}

It is expected that the surface bias of the nearside jet will cause the recoil parton to travel through a significant amount of the medium, therefore making the awayside the ideal place to study any parton energy loss and jet modification.  The awayside Gaussian widths (Figure \ref{fig:asw}) indicate significant broadening of the jet peak in Au--Au compared to p--p.  The awayside $I_{AA}$ (Figure \ref{fig:asiaa}) indicates significant softening of the awayside jets in Au--Au.  This modification can also be clearly seen in $D_{AA}$ (Figure \ref{fig:asdaa}).  Since $\Delta B \sim 1 - 2\mbox{ GeV}/c$, it is clear that most (but not all) of the high-$p_T^{assoc}$ suppression is compensated for by low-$p_T^{assoc}$ enhancement.  All of these observations are indicative of significant jet quenching on the awayside.  The observed modifications in the jet shapes and yields between Au--Au and p--p are in qualitative agreement with the radiative energy loss picture while the black-and-white models are disfavored.  

\section{``2+1'' and Jet-Hadron Correlations}
\begin{figure}
\begin{center}
\includegraphics[width=160mm]{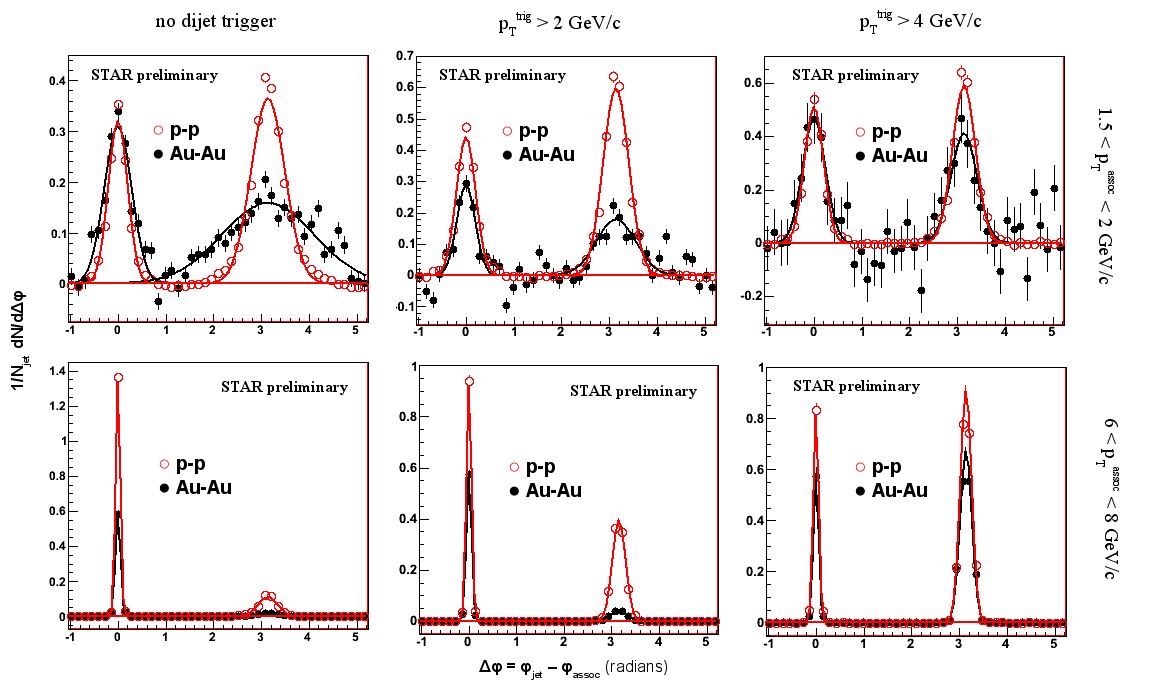}%\hspace{1pc}%
\caption{\label{fig:2p1}Jet-hadron correlations with the requirement of a dijet trigger (high-$p_T$ hadron opposite the reconstructed jet axis, $|\varphi_{jet} - \varphi_{trig} - \pi| < 0.2$): left $p_T^{trig} > 0\mbox{ GeV}/c$, center $p_T^{trig} > 2\mbox{ GeV}/c$, right $p_T^{trig} > 4\mbox{ GeV}/c$.  The reconstructed jet energy is 10 \textless $p_T^{jet}$ \textless 20 GeV/$c$ and two $p_T^{assoc}$ bins are shown: top $1.5 < p_T^{assoc} < 2\mbox{ GeV}/c$ and bottom $6 < p_T^{assoc} < 8\mbox{ GeV}/c$.  As the dijet trigger $p_T$ threshold rises the awayside jet peak in Au--Au (black closed symbols) becomes narrower and taller to match the awayside peak in p--p (red open symbols).}
\end{center}
\end{figure}

In 2+1 correlations, a pair of back-to-back high-$p_T$ hadrons is used as a dijet proxy.  Typical dihadron correlations are then performed with respect to both trigger particles\cite{ref:2p1}.  It is suspected that the dijet proxy requirement in 2+1 correlations selects jets which are unmodified, possibly due to surface bias.  The systematics of 2+1 correlations can be explored through jet-hadron correlations, by performing the jet-hadron analysis when there is a high-$p_T$ hadron opposite the reconstructed trigger jet.  Furthermore, the 2+1 correlation analysis used a $p_T^{assoc}$ cut-off at 1.5 GeV/$c$; in jet-hadron correlations the analysis can be extended below this cut.  In the jet-hadron correlations shown in Figure \ref{fig:2p1} a dijet trigger is required in the range $|\varphi_{jet} - \varphi_{trig} - \pi| < 0.2$.  As the $p_T$ cut on the dijet trigger is raised from 0 to 2 to 4 GeV/$c$, the awayside jet peak in Au--Au becomes more and more similar to p--p.  Additionally, a comparison of the jet and event populations selected by the various $p_T^{trig}$ thresholds indicates that 2+1 correlations are biased towards harder jets as well as more peripheral events.  This supports the idea that the jet population selected in 2+1 correlations is biased towards unmodified jets.  

\section{Jets with respect to the Event Plane}
Future work on jet-hadron correlations will include an investigation of the pathlength dependence of partonic energy loss via an analysis of the awayside jet modification as a function of the relative angle between the trigger jet and the event plane.  Before performing the jet-hadron analysis with respect to the event plane, however, it is necessary to first investigate the relation between the trigger jet and the reaction plane.  

\begin{figure}
\includegraphics[width=90mm]{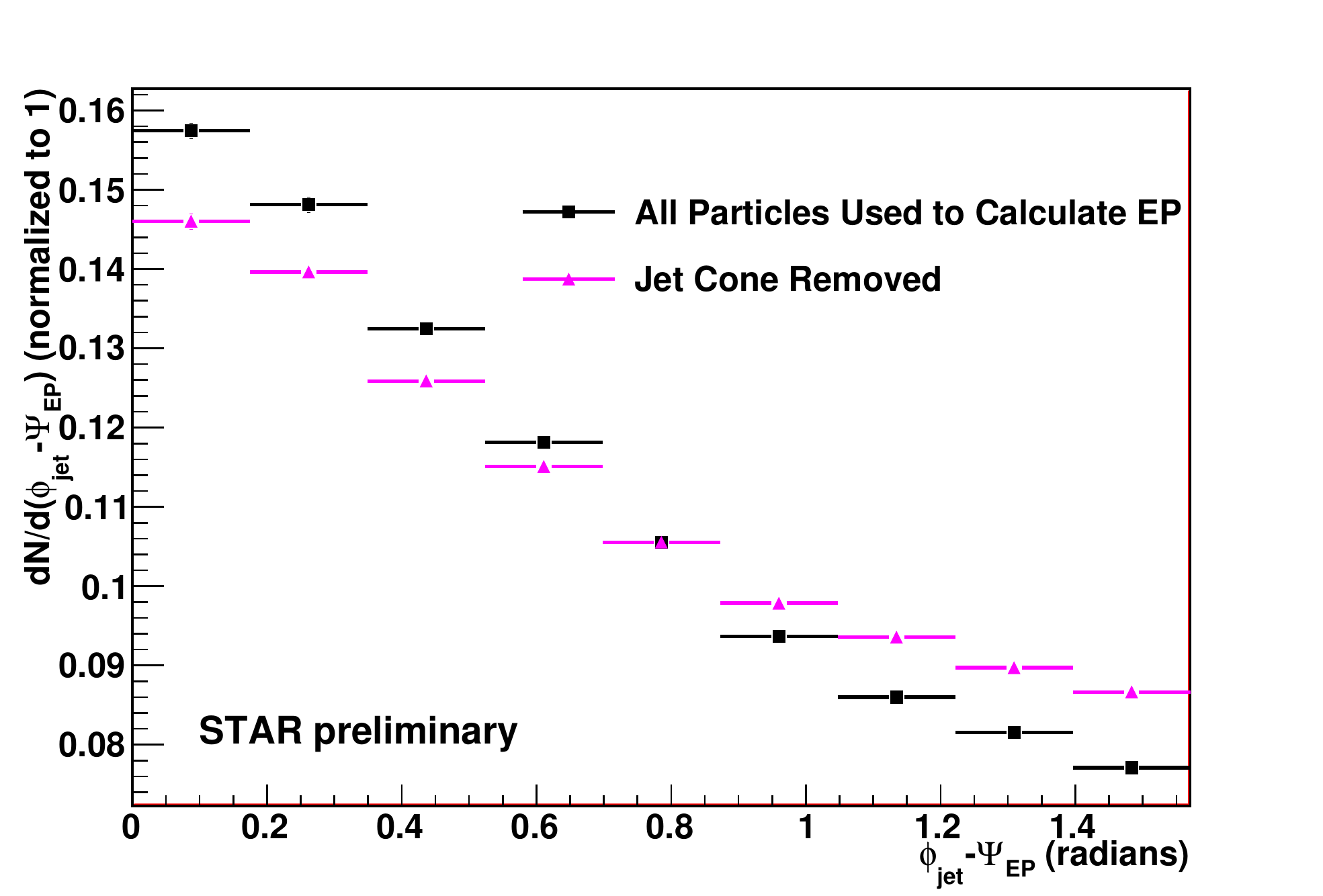}%\hspace{1pc}%
\begin{minipage}[b]{14pc}\caption{\label{fig:jEP}The relative angle between a HT trigger jet ($p_T^{jet} > 10\mbox{ GeV}/c$) and the event plane.  The event plane is reconstructed from all charged tracks in the TPC with $p_T < 2\mbox{ GeV}/c$ and without $p_T$ weighting using Equation \ref{eq:EP}.  The result (black squares) is compared to the correlation between the jet axis and event plane when tracks within a cone of $R = 0.4$ in $(\eta,\varphi)$-space around the reconstructed jet axis are excluded from the event plane calculation (pink triangles).  }
\end{minipage}
\end{figure}

\begin{figure}
\includegraphics[angle=90,width=90mm]{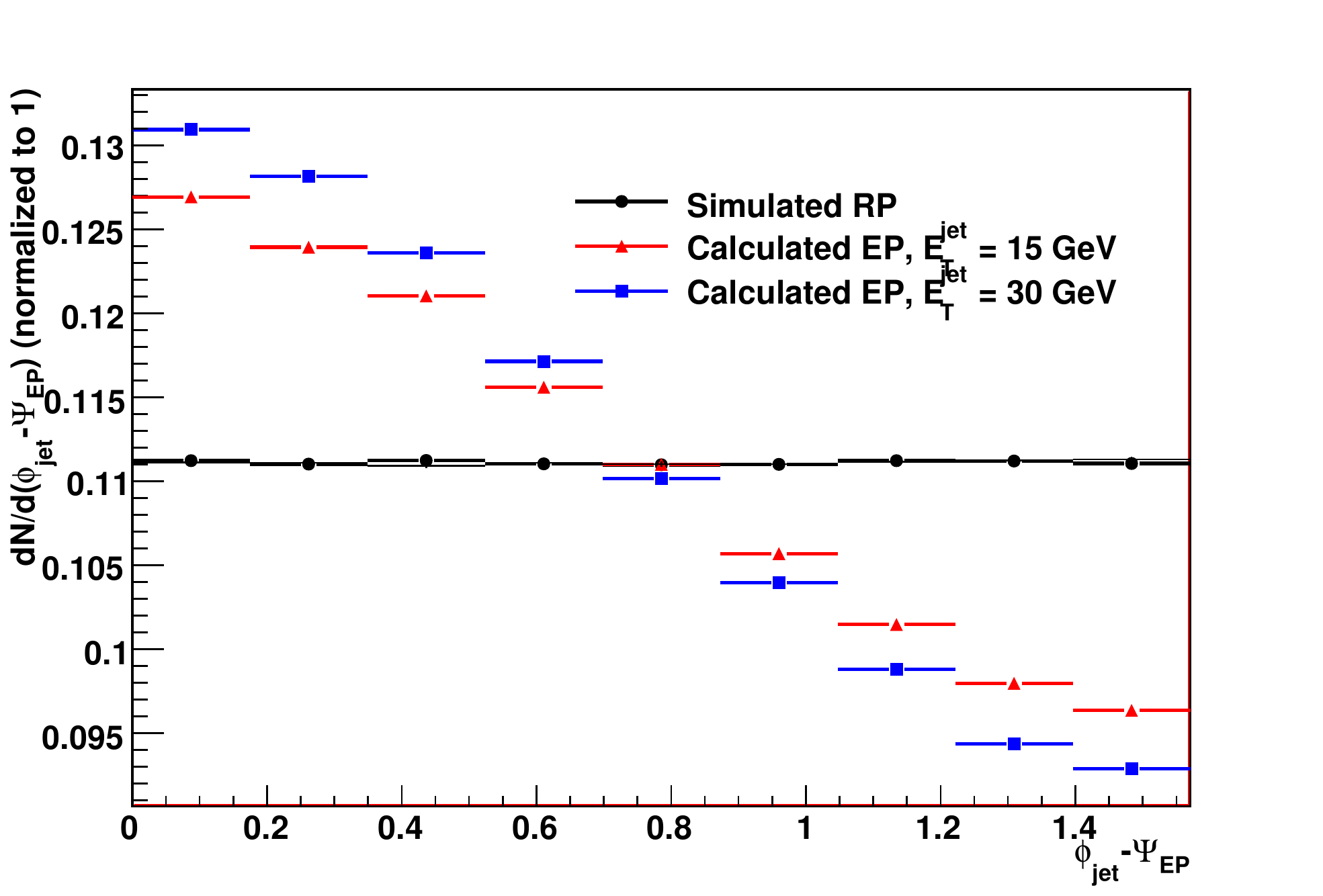}%\hspace{1pc}%
\begin{minipage}[b]{14pc}\caption{\label{fig:jEPsim}A jet is embedded in a thermal background ($T = 0.291$ GeV) with $p_T$- and centrality-dependent $v_2$ (10-20\% centrality).  By construction $v_2^{jet} = 0$ (black circles).  The correlation between the jet axis and the reconstructed event plane is shown for PYTHIA jets with $E_T^{PYTHIA} = 15$ GeV (red triangles) and $E_T^{PYTHIA} = 30$ GeV (blue squares).}
\end{minipage}
\end{figure}

Figure \ref{fig:jEP} shows that there is a significant correlation between the axis of a HT trigger jet and the event plane (reconstructed using Equation \ref{eq:EP}).  
\begin{equation}\label{eq:EP}
\tan(2\Psi_{EP}) = \frac{\sum_n \sin(\varphi_n)}{\sum_n \cos(\varphi_n)}
\end{equation}
Although this result appears to clearly show a non-zero jet $v_2$, there is another potential competing effect: the jet may be biasing the event plane calculation.  When tracks within a cone of $R = 0.4$ in $(\eta,\varphi)$-space around the reconstructed jet axis are excluded from the event plane calculation, the correlation between the reconstructed jet and the event plane is reduced.  Therefore the presence of a jet affects the calculation of the event plane, but the question remains: is there a jet $v_2$ beneath the artificial jet-event plane bias?  The jet-event plane bias can be further investigated through simulation: PYTHIA\cite{ref:pythia}\cite{ref:pythia2} jets are embedded isotropically in a thermal background.  The jets are randomly distributed with respect to the reaction planes of the background event and therefore entirely uncorrelated ($v_2^{jet} = 0$ by construction).  The event plane is calculated in the standard way (Equation \ref{eq:EP}) and the distribution of the relative angle between the jet and the event plane is shown in Figure \ref{fig:jEPsim}.  For PYTHIA jets with $E_T^{PYTHIA} = 15$ GeV the jet has an artificial $v_2^{obs}$ of $\sim7\%$ (before dividing by the event plane resolution, meaning the artificial jet $v_2$ is actually higher), and for jets with $E_T^{PYTHIA} = 30$ GeV the artificial jet $v_2^{obs}$ is $\sim9\%$.  Therefore the jet pulls the event plane significantly.  Attempts to remove the jet-event plane bias in simulation (by removing particles in the jet cone or in a $\eta$ wedge, or by using the Lee-Yang Zeros method to calculate the event plane\cite{ref:LYZ}) have been so far unsuccessful.  Future work will investigate the possibility of using forward detectors (FTPC, BBC, ZDC-SMD) to calculate the event plane in the presence of a jet, but long-range correlations associated with the jet may have an effect on event plane reconstruction at forward rapidity as well.  

\section{Conclusions}
Jet-hadron correlations can be used to investigate the modification of partons that traverse the medium created in heavy ion collisions.  The results indicate that in high tower trigger events the recoil jet is significantly broadened and softened in Au--Au compared to p--p, in qualitative agreement with the radiative energy loss picture.  Furthermore, most of the high-$p_T$ suppression is balanced by low-$p_T$ enhancement.  Jet-hadron correlations also indicate that the dijet trigger requirement in 2+1 correlations biases the jet sample towards unmodified jets.  A study of jets with respect to the event plane shows that the presence of a jet significantly biases the event plane calculation.  An analysis of jet $v_2$ and jet-hadron correlations with respect to the event plane is in progress.  

\section*{Acknowledgments}
This  research  was  supported  in  part  by  an  award from  the  Department  of  Energy  (DOE)  Office  of  Science  Graduate  Fellowship  Program administered  by  the  Oak  Ridge  Institute  for  Science  and  Education  for  the  DOE.  This work was also supported in part by the facilities and staff of the Yale University Faculty of Arts and Sciences High Performance Computing Center.

\bibliography{references}{}
\end{document}